\documentclass[a4paper,11pt]{article}
\pdfoutput=1 

\usepackage{arXivpub} 
\usepackage{graphicx}
\usepackage{subcaption}
\usepackage{multirow}
\usepackage{lineno}

\title{\boldmath Test beam measurement of the first prototype of the fast silicon pixel monolithic detector for the TT-PET project }

\author[a,1]{L. Paolozzi,\note{Corresponding author.}}
\author[b]{Y. Bandi,}
\author[a]{M. Benoit,}
\author[c]{R. Cardarelli,}
\author[a]{S. D\'ebieux,}
\author[b]{D. Forshaw,}
\author[a,d]{D. Hayakawa,}
\author[a]{G. Iacobucci,}
\author[e]{M. Kaynak,}
\author[b]{A. Miucci,}
\author[a,f]{M. Nessi,}
\author[d]{O. Ratib,}
\author[a,d]{E. Ripiccini,}
\author[e]{H. R\"ucker,}
\author[a]{P. Valerio,}
\author[b]{and M. Weber}

\affiliation[a]{University of Geneva,\\Rue du G\'en\'eral-Dufour 24, Geneva, Switzerland}
\affiliation[b]{University of Bern,\\Sidlerstrasse 5, Bern, Switzerland}
\affiliation[c]{INFN Section of Roma Tor Vergata,\\Via della ricerca scientfica 1, Roma, Italy}
\affiliation[d]{Institute of Translational Molecular Imaging (ITMI), University of Geneva,\\Geneva, Switzerland}
\affiliation[e]{IHP,\\Im Technologiepark 25, Frankfurt (Oder), Germany}
\affiliation[f]{CERN,\\Geneva, Switzerland}

\emailAdd{lorenzo.paolozzi@unige.ch}

\abstract{The TT-PET collaboration is developing a PET scanner for small animals with $ 30 ~\mathrm{ps} $ time-of-flight resolution and sub-millimetre 3D detection granularity. The sensitive element of the scanner is a monolithic silicon pixel detector based on state-of-the-art SiGe BiCMOS technology.  The first ASIC prototype for the TT-PET was produced and tested in the laboratory and with minimum ionizing particles. The electronics exhibit an equivalent noise charge below $ 600 ~\mathrm{e^{-}} ~\mathrm{RMS} $ and a pulse rise time of less than $ 2 ~\mathrm{ns} $, in accordance with the simulations. The pixels with a capacitance of $ 0.8 ~\mathrm{pF} $ were measured to have a detection efficiency greater than $ 99 \% $ and, although in the absence of the post-processing, a time resolution of approximately $ 200 ~\mathrm{ps} $. }


\begin{document}
\maketitle
\flushbottom

\section{Introduction}
\label{sec:intro}

\begin{figure}[htpb]
\begin{subfigure}{0.49\textwidth}
\centering
\includegraphics[width = \textwidth]{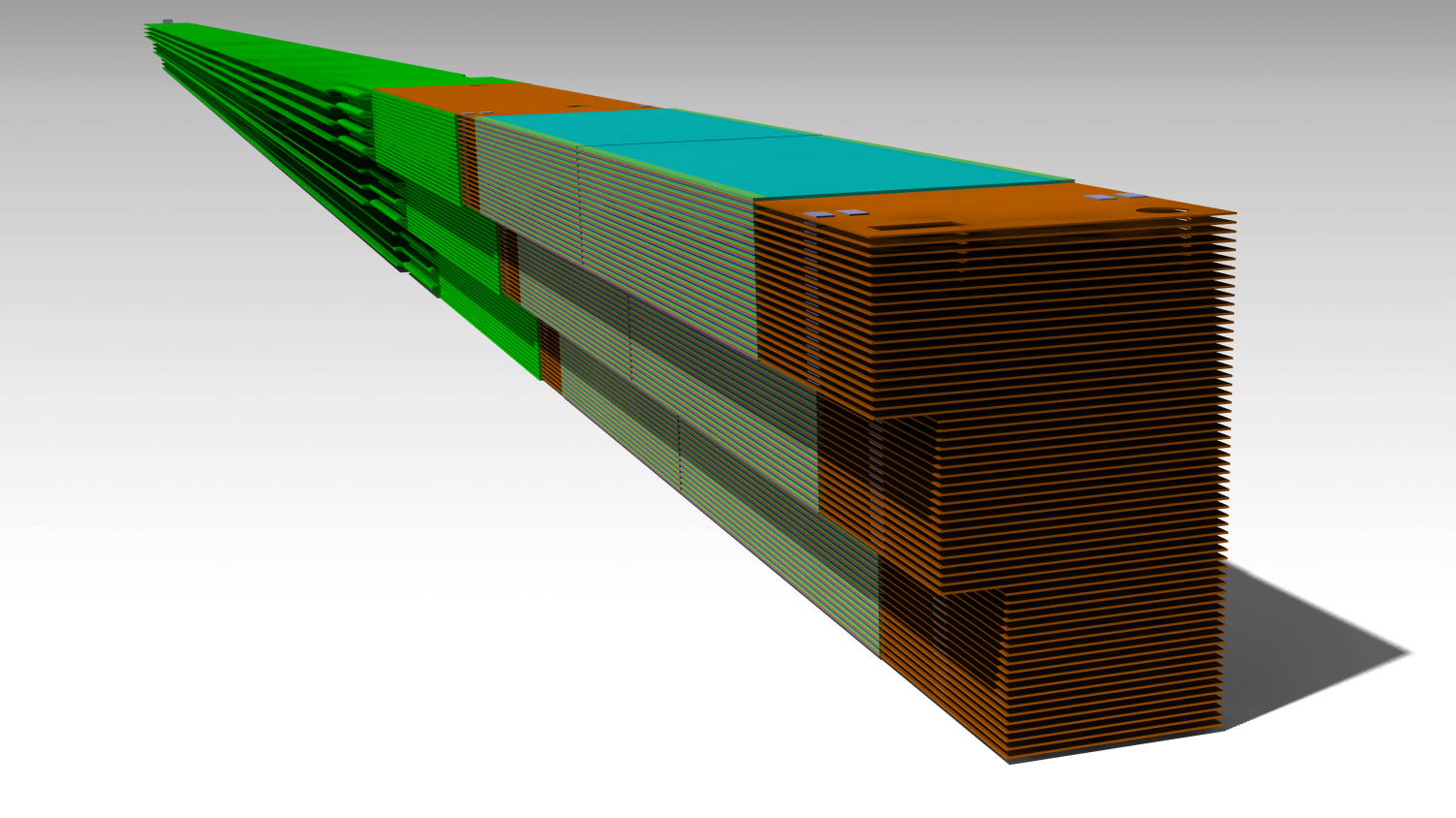}
\end{subfigure}
\begin{subfigure}{0.49\textwidth}
\centering
\includegraphics[width = 0.85 \textwidth]{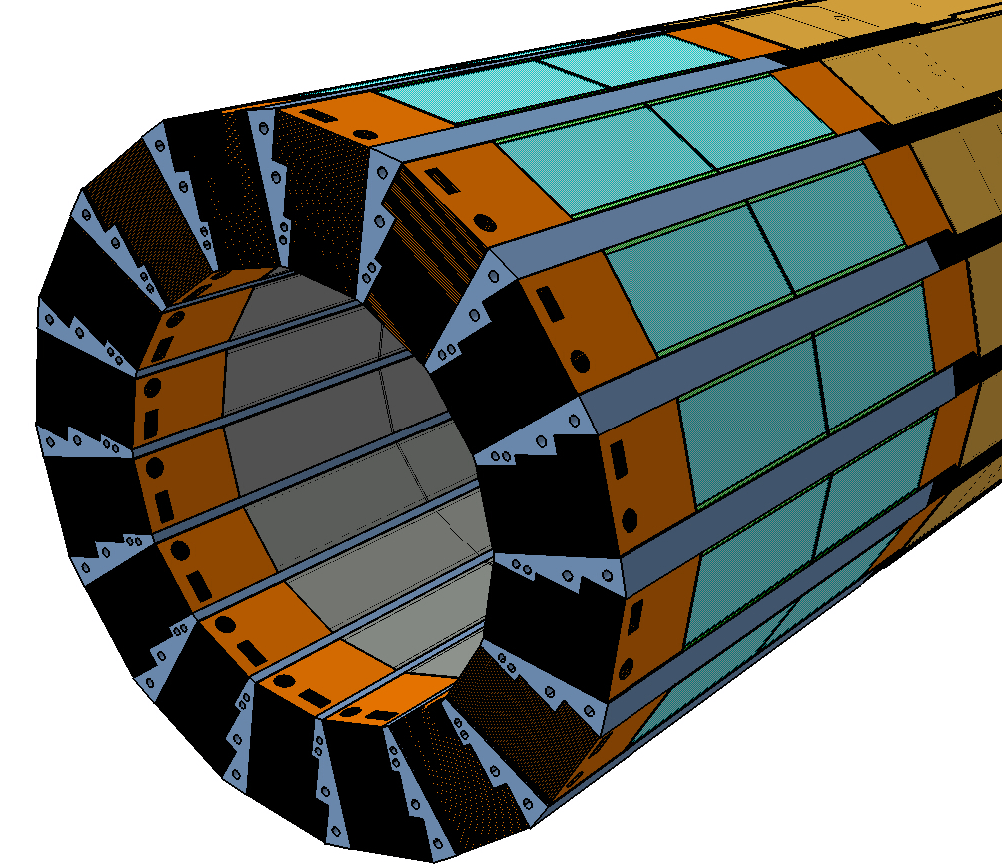}
\end{subfigure}
\caption{\label{fig:scanner} CAD model of a single wedge (left) and the entire TT-PET scanner (right).}
\end{figure}

The feasibility of a silicon pin detector with $ 100 ~\mathrm{ps} $ time resolution for Minimum Ionizing Particles (MIPs) with $ 1 ~\mathrm{pF} $ pixel capacitance was demonstrated in a previous work exploiting a parallel plate readout and a fast low-noise amplifier based on SiGe HBT technology \cite{100ps}. That result, obtained for a $ 100 ~\mathrm{\mu m} $ thick sensor, ensured that, with the proper sensor geometry and a fast charge-integrating amplifier, the detector time resolution can be expressed as \cite{100ps,MLCD,Cartiglia}: 

\begin{equation}
\label{eq:timeres}
\sigma_{t} =~ \frac{Rise ~time}{\frac{Signal}{Noise}} =~  \frac{ENC}{\frac{dQ}{dx}v_d} ~,
\end{equation}
where $ \frac{dQ}{dx} $ is the primary charge produced per unit length in the sensor, $ v_d $ is the average drift velocity of the charge carriers and $ ENC $ is the equivalent noise charge of the preamplifier. The formula does not take into account the irreducible sensor contribution that comes from  the charge collection noise (or Landau noise) \cite{phdPaolozzi}, which contributes approximately $ 30 ~\mathrm{ps} ~\mathrm{RMS} $ for a $ 100 ~\mathrm{\mu m} $ thick pin diode \cite{phdPaolozzi,sitimemodel}. 
That work was the starting point for the development of the monolithic pixel detector for the Thin-Tof PET (TT-PET) project \cite{grant,PSD11}, which aims to develop a time-of-flight positron emission tomography scanner for small animals based on silicon pixel sensors. The scanner is expected to achieve $ 30~\mathrm{ps} $ time resolution for $ 511 ~\mathrm{keV} $ photons, equivalent to $ 100 ~\mathrm{ps} $ time resolution for MIPs (\cite{Emanuele} and eq. \ref{eq:timeres}). The basic TT-PET photon detection element is made by a $ 50 ~\mathrm{\mu m} $ thick lead converter, a $ 100 ~\mathrm{\mu m} $ thick monolithic silicon pixel sensor and a $ 50 ~\mathrm{\mu m} $ thick polyimide flex circuit. The scanner is segmented into 16 equally sized wedge-like structures which are formed by stacking the basic detection element vertically on top of each other 60 times (Figure \ref{fig:scanner}). The system will have approximately 1.5 million pixels, with a readout granularity of $ 500 \times 500 \times 200 ~\mathrm{\mu m}^3 $. 

The specifications of the TT-PET chip in terms of power budget, time resolution and read-out channel density required the integration of both the sensor matrix and the front-end electronics in a monolithic structure, using state-of-the-art SiGe BiCMOS technology. A detector with a very high 3D detection granularity necessitates low power consumption and a minimization of the sensor interconnections. The specifications of the analogue components of the final ASIC are shown in Table \ref{tab:specs}. 

\begin{table}[htbp]
\centering
\caption{\label{tab:specs} Specifications of the monolithic pixel detector chip for the TT-PET scanner.}
\smallskip
\begin{tabular}{|l|c|}
\hline
ASIC lenght & $ 24 ~\mathrm{mm} $\\
ASIC width & $ 7, ~9, ~11 ~\mathrm{mm} $\\
Pixel Size & $ 500 \times 500 ~\mathrm{\mu m^2} $\\
Pixel Capacitance (comprised routing) & $ <750 ~\mathrm{fF} $\\
Power Density Budget & $ 40 ~\mathrm{mW/cm^2} $\\
Preamplifier $ ENC $ RMS ($ 750 ~fF $ input capacitance) & $ 600 ~\mathrm{e^{-}} $\\
Preamplifier Rise Time ( $ 10\% -90 \% $ ) & $ 800 ~\mathrm{ps} $\\
Time Resolution RMS for MIPs & $ 100 ~\mathrm{ps} $\\
\hline
\end{tabular}
\end{table}

The chip will be produced on a $ 1 ~\mathrm{k \Omega cm} $ resistivity substrate to achieve full bulk depletion and to avoid the breakdown in the inter-pixel region at high voltages. In order to guarantee the uniformity of both the electric field and the weighting field, it will be necessary to thin down to $ 100 ~\mu m $ and back-side metallize the chip. The electric field in the bulk will be of $ 2\div3 ~\mathrm{V/\mu m} $, in order to saturate the drift velocity of the charge carriers in silicon. 

\section{Design, simulation and characterization of the first ASIC prototype}
\label{sec:chip}

\begin{figure}[htpb]
\begin{center}
\includegraphics[width = 0.9\textwidth]{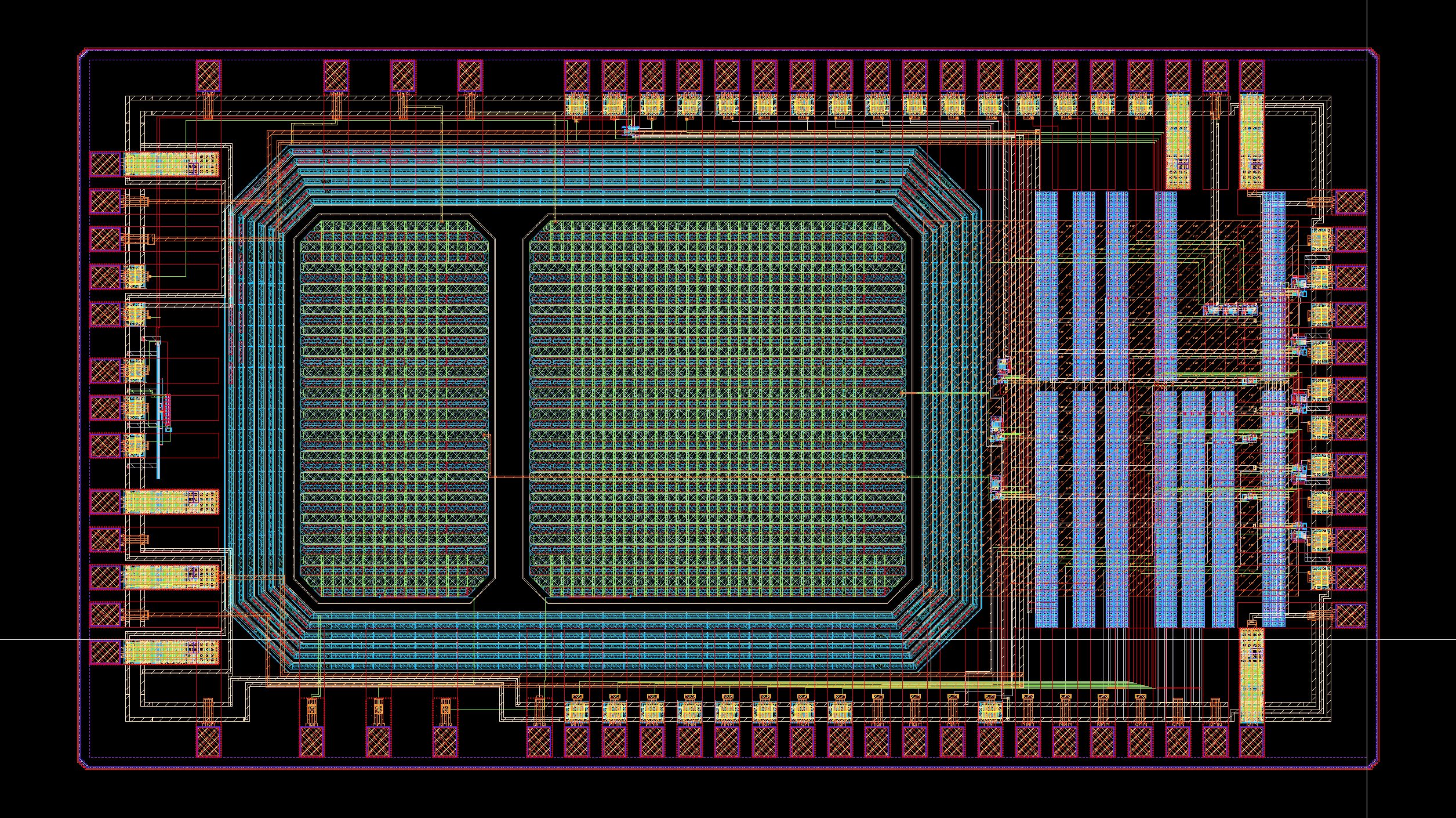}
\end{center}
\caption{\label{fig:layout} Layout of the first monolithic test chip of the TT-PET project. The guard ring separates the sensitive volume from the front-end electronics.}
\end{figure}

The first monolithic test chip for the TT-PET project was specifically designed to prove the functionality and performance of the front-end electronics on a very high resistivity substrate and verify the high voltage tolerance of the full custom guard ring and inter pixel structures implemented for the first time in a SiGe BiCMOS process. The design was done using the SG13S technology \cite{sg13s} from IHP microelectronics, which provides a fast, low-noise, low-current SiGe HBT for the pre-amplification stage. The production was done using a multi-project wafer via Europractice-IC services\footnote{\url{http://www.europractice-ic.com/}}. The chip layout is shown in Figure \ref{fig:layout}.

The chip contains a small $ \left( 900 \times 450 ~\mathrm{\mu m}^2 \right) $ and a large $ \left( 900 \times 900 ~ \mu m^2 \right) $ n-in-p pixel inside a common guard ring, where the pixels are spaced by $ 100 ~\mathrm{\mu m} $. The electronics, placed outside the guard ring, consist of a SiGe-HBT preamplifier and a discriminator with time-over-threshold measurement capability. The chip was produced using a wafer of $ 1 ~\mathrm{k \Omega cm} $ resistivity. It was not thinned, nor had backside metallization when being characterised at test beam. This prevented the saturation of the drift velocity of the charge carriers and worsened the uniformity of the weighting field for the signal induction. A positive bias voltage was applied directly to each pixel using poly-silicon biasing resistors. 
 
\begin{figure}[htpb]
\includegraphics[width = \textwidth]{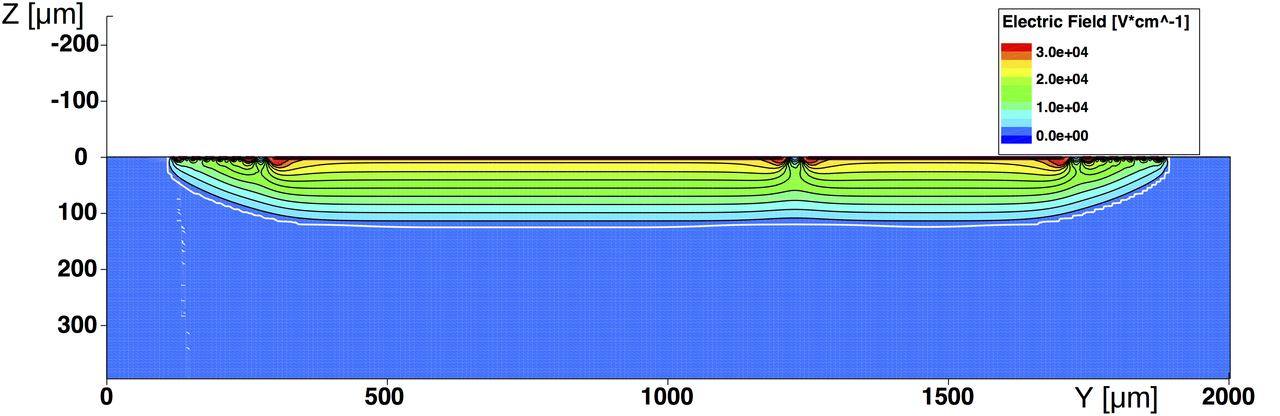}
\caption{\label{fig:TCAD} Results of the TCAD simulation: cross section of the sensor showing the depletion region and the electric field intensity underneath the small pixel, large pixel and guard ring. The white line represents the limit of the depletion region. The fact that the thick substrate was not thinned to $ 100 ~\mathrm{\mu m} $ and the absence of the backplane metallization cause an electric field gradient as a function of the sensor depth and prevent the saturation of the drift velocity of the charge carriers.}
\end{figure}

Figure \ref{fig:TCAD} shows the TCAD\footnote{\url{http://www.synopsys.com} - Software: Synopsys Sentaurus} simulation of the sensor depletion profile in the direction crossing both pixels for a bias voltage of $ 160 ~\mathrm{V} $. The simulation shows that the depletion depth is approximately $ 130 ~\mathrm{\mu m} $ and is not uniform underneath the pixel area. The electric field, well below the desired $ 2\div3 ~\mathrm{V/\mu m} $, is not uniform in depth and is too low to saturate the charge-carriers drift velocities. Both effects are expected to degrade the performance of the un-thinned sensor that we operated in the test beam.

\begin{figure}[htpb]
\begin{center}
\includegraphics[width = 0.45\textwidth]{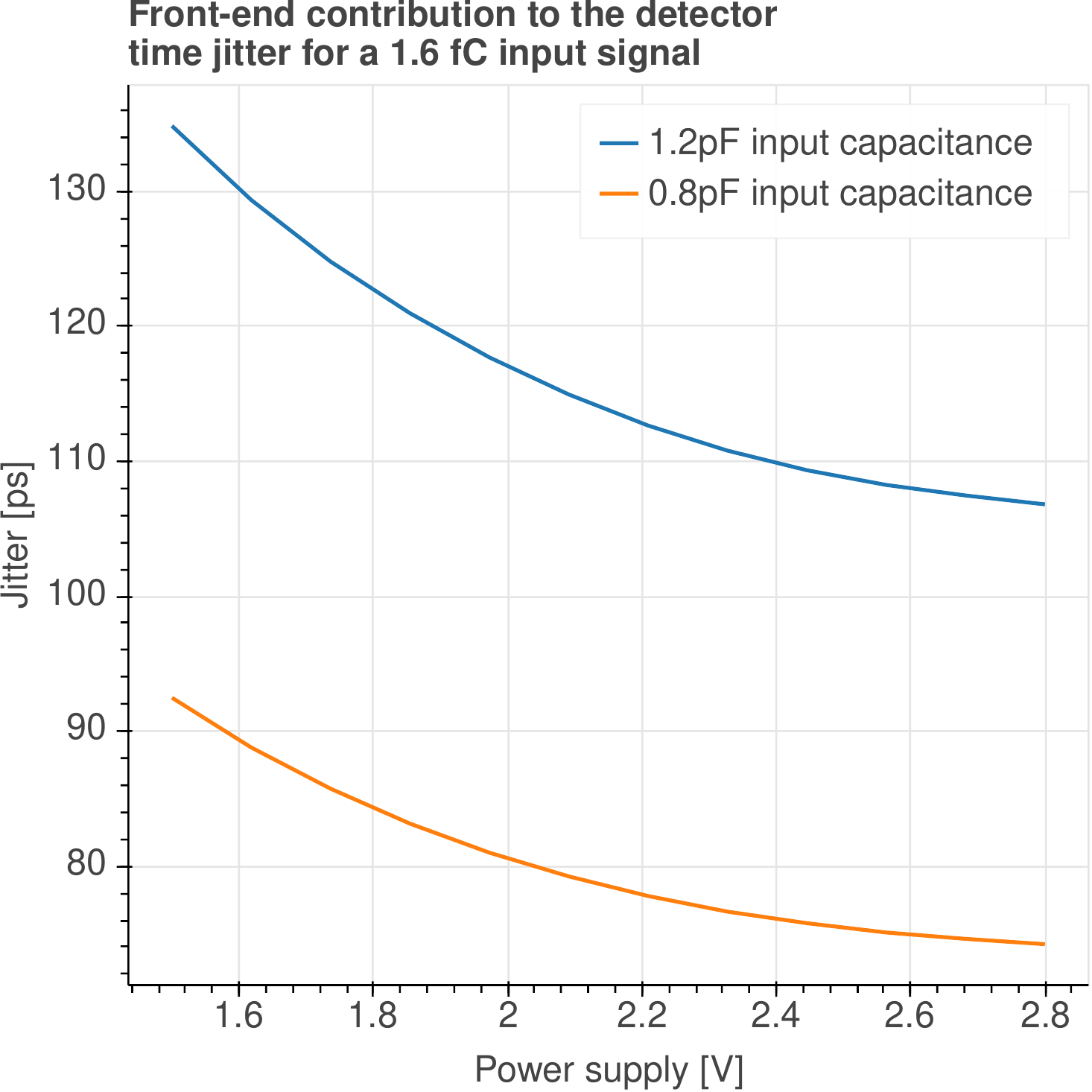}
\end{center}

\caption{\label{fig:TJPC} Results of the Cadence simulation: the front-end electronics contribution to the time jitter for a $ 1.6 ~\mathrm{fC} $ input signal as a function of the preamplifier power supply, for the capacitance values of the two pixels under study. This is expected to be the dominating contribution to the time jitter in the parallel-plate approximation with saturated drift velocity of the charge carriers.}
\end{figure}

The capacitance of the small and large pixels have been measured at the same operating point of $ 160 ~\mathrm{V} $ to be $ 0.8 ~\mathrm{pF} $ and  $ 1.2 ~\mathrm{pF} $, respectively. For these capacitance values, the expected preamplifier noise ($ ENC $) from Cadence\footnote{\url{https://www.cadence.com} - Software: Virtuoso ADL} simulation is less than $ 600 ~\mathrm{e^{-}} ~\mathrm{RMS} $ for the small pixel and $ 750 ~\mathrm{e^{-}} ~\mathrm{RMS} $ for the large one. The pulse rise time at the output of the preamplifier is equivalent to the charge collection time. The working voltage is adjustable and has been optimized using Cadence in terms of the contribution to the time jitter and the power consumption (Figure \ref{fig:TJPC}). The breakdown voltage was measured to be approximately $ 165 ~\mathrm{V} $.

\section{Experimental setup at the CERN SPS test beam}
\label{sec:setup}

The performance of the monolithic test chip has been measured at the SPS test beam line at CERN with a beam of $ 180 ~\mathrm{GeV/c} $ momentum pions. Two chips were glued on two identical test boards with conductive glue, where the substrate was referenced to ground. The discriminator threshold was set individually for each readout channel. The discriminator outputs were transmitted using coaxial lines with $ 50 ~\mathrm{\Omega} $ impedance.

\begin{figure}[htpb]
\includegraphics[width = \textwidth]{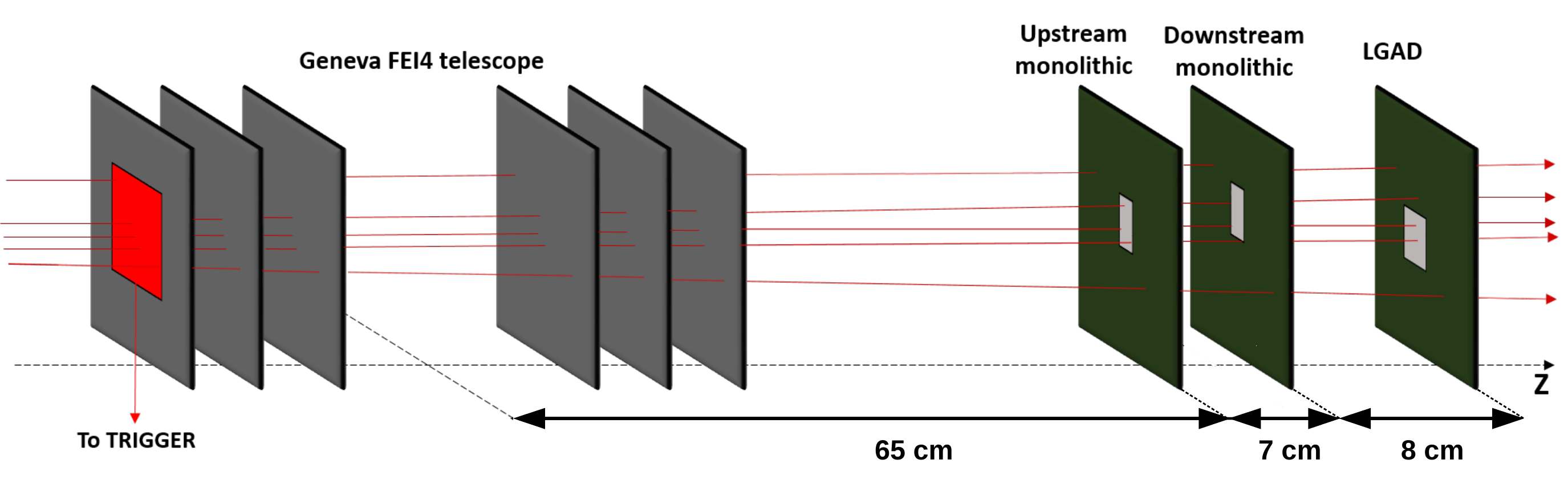}

\caption{\label{fig:setup} Schematic representation of the experimental setup showing, from left to right, the six planes of the telescope, the upstream and downstream monolithic test chips, and the LGAD. The horizontal red arrows represent the pion beam. The telescope trigger area is larger than the senstitive area of the detectors under test.}
\end{figure}

A $ 50 ~\mathrm{\mu m} $ thick LGAD sensor \cite{UFSD} with $ 30 ~\mathrm{ps} $ time resolution produced by CNM \cite{Sebastian}, placed on a third board and readout by the same electronics of \cite{100ps}, was used as a time reference for the measurements. The signals from the monolithic sensors and the LGAD were sent to an oscilloscope with $ 1 ~\mathrm{GHz} $ analogue bandwidth. The three boards were placed downstream with respect to the beam of the Geneva FE-I4 telescope \cite{telescope}, which provided the trigger to the oscilloscope and the trajectory of the pions. Figure \ref{fig:setup} shows a schematic representation of the setup. 

All the data presented on this paper were acquired with a bias voltage of $ 160 ~\mathrm{V} $ applied to the monolithic pixels and $ 230 ~\mathrm{V} $ applied to the LGAD. The amplifier power consumption was $ 350 ~\mathrm{\mu W/channel} $, corresponding to a simulated noise ($ ENC $) of $ 750 ~\mathrm{e^{-}} ~\mathrm{RMS} $ for the large pixel and $ 550 ~\mathrm{e^{-}} ~\mathrm{RMS} $ for the small pixel. The discriminator signals from the small pixel of the upstream sensor, the large pixel from the upstream sensor, the small pixel of the downstream sensor, and the LGAD were recorded. 

A total of $ 1\,104\,396 $ events were triggered by the telescope in an area of $ 3200 \times 2900 ~\mathrm{\mu m}^2 $. Due to the position of the detector under test, which was not in the center of the telescope, a selection of the tracks with a slope within two standard deviations from the mean value of their distribution was used in order to reduce the effect of multiple scattering. After this selection, the dataset consisted of $ 972\,046 $ events in the triggered region.

\section{Test beam results}
\label{sec:results}

\subsection{Efficiency and uniformity of response}
\label{sec:efficiency}

\begin{figure}[htpb]
\begin{subfigure}{0.49\textwidth}
\centering
\includegraphics[width = \textwidth]{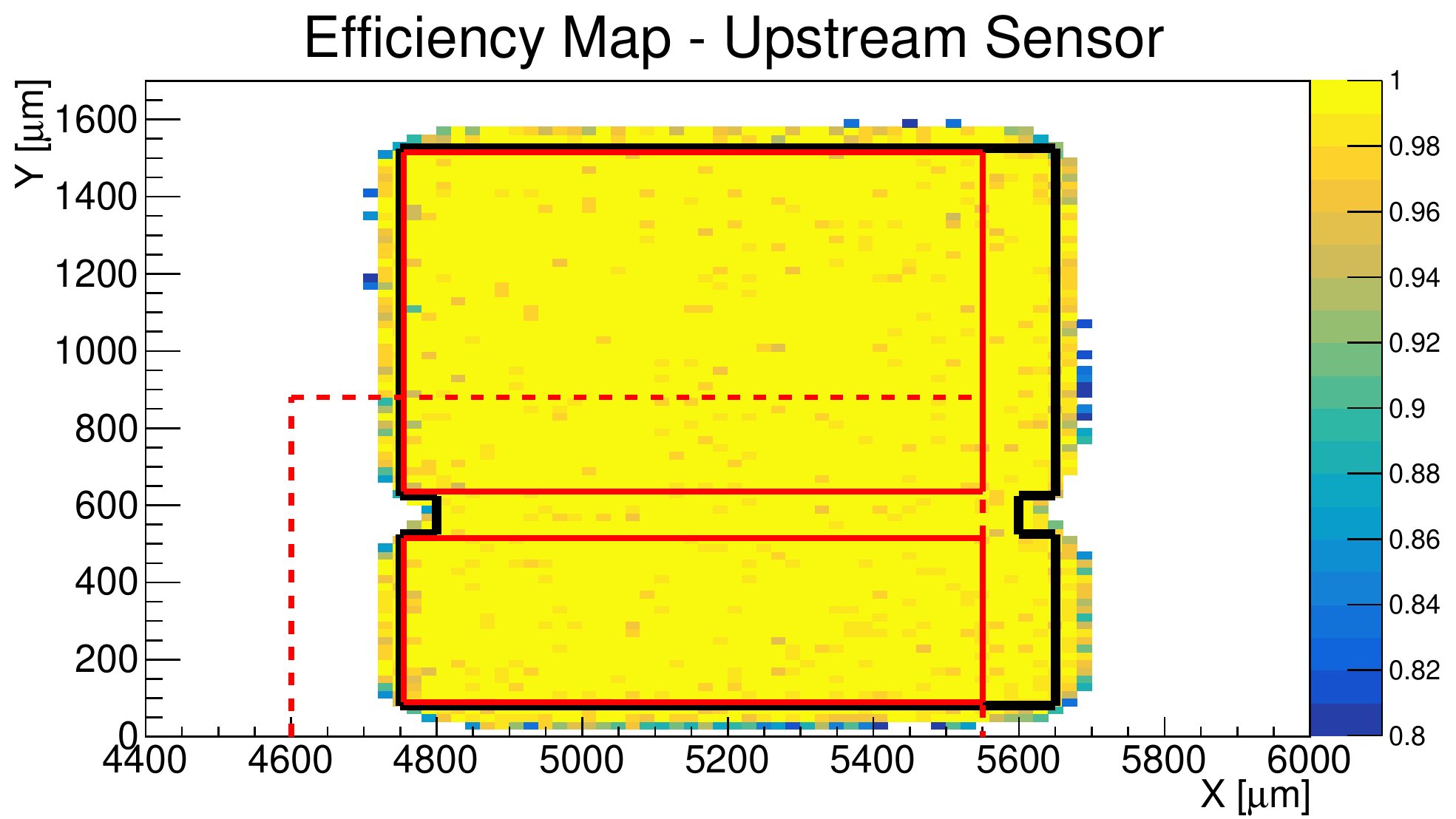}
\end{subfigure}
\begin{subfigure}{0.49\textwidth}
\includegraphics[width = \textwidth]{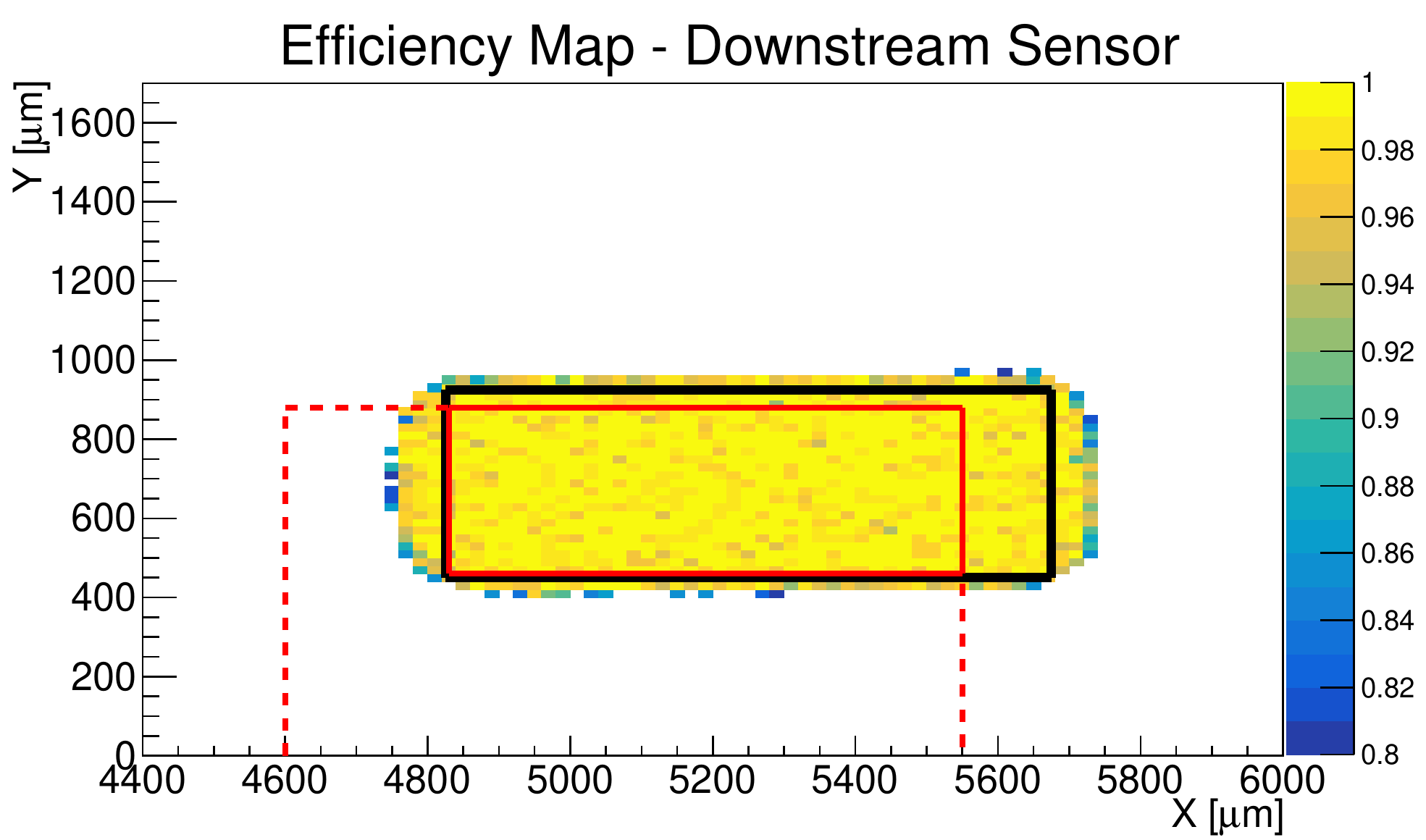}
\end{subfigure}
\caption{\label{fig:EFFMAP} Efficiency map of the upstream chip (left) and the downstream small pixel (right). The areas contoured in black represent the regions of the monolithic pixels selected to measure the efficiency; the dashed rectangles represent the projection of the LGAD active area on the monolithic chips. The red rectangles are the regions of interest used for the time resolution measurement. In the downstream sensor, only the small pixel was read out.}
\end{figure}

The efficiency map for the upstream sensor and the downstream small pixel are shown in Figure \ref{fig:EFFMAP}. The efficiency is calculated by selecting the tracks pointing to a region of the detector under test and calculating the fraction of events detected by the sensor. The three pixels show a high and uniform efficiency. The areas contoured in black in Figure \ref{fig:EFFMAP} correspond to the region of the pixels selected to study the efficiency. The dashed rectangle is the projection of the LGAD active area on the sensors. The region of interest on which the time resolution has been studied is represented by the red rectangles.

The measured efficiency for MIP detection is $ (99.79 \pm 0.01) \%$ for the upstream sensor and $ (99.09 \pm 0.04) \%$ for the small pixel of the downstream sensor. The smaller value obtained for the downstream pixel is probably due to the multiple scattering produced by the upstream board, which is reducing the tracking precision of the telescope. In order to reduce this effect, the efficiency of the downstream pixel has been measured in the same region selecting only the events from the telescope that were detected also by the LGAD. The downstream small pixel efficiency with this selection was measured to be $ (99.88 \pm 0.04) \% $.

\begin{figure}[htpb]
\centering
\includegraphics[width = 0.55\textwidth]{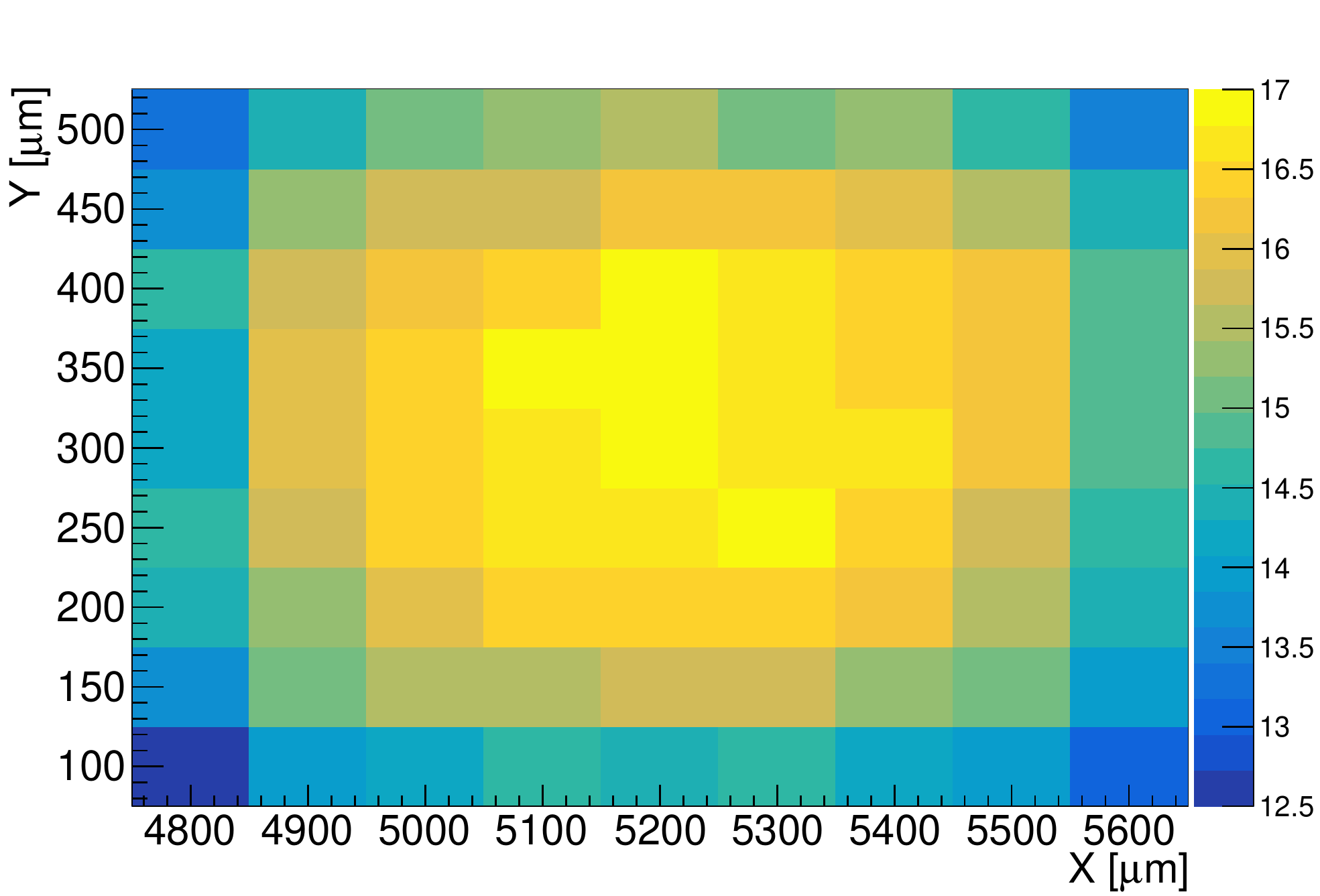}
\caption{\label{fig:MAPTOTUS} Map of the most probable time-over-threshold values for the upstream small pixel. For small signals, the time-over-threshold at the output of the amplifier is proportional to the charge deposited in the pixel.}
\end{figure}

The uniformity of response over the pixel area was also studied. The most probable time-over-threshold (ToT) of the signal was used to estimate the charge deposited in the different regions of the depleted sensor volume. Figure \ref{fig:MAPTOTUS} shows the map of the most probable ToT on the small pixel from the upstream chip. The average charge is measured to be smaller in the region close to the pixel edge, with $ \sim 15\% $ less charge compared to the center of the pixel. As shown before, the TCAD simulation presented in Figure \ref{fig:TCAD} indicates that the depletion depth was expected to be constant in the selected area. We conclude from the ToT measurement that the simulation is not able to reproduce the effect, probably because it does not contain the models for the charge sharing and the weighting field for fast signals when the reference ground is made by high-resistivity silicon.

\begin{figure}[htpb]
\begin{subfigure}{0.49\textwidth}
\centering
\includegraphics[width = 0.92\textwidth]{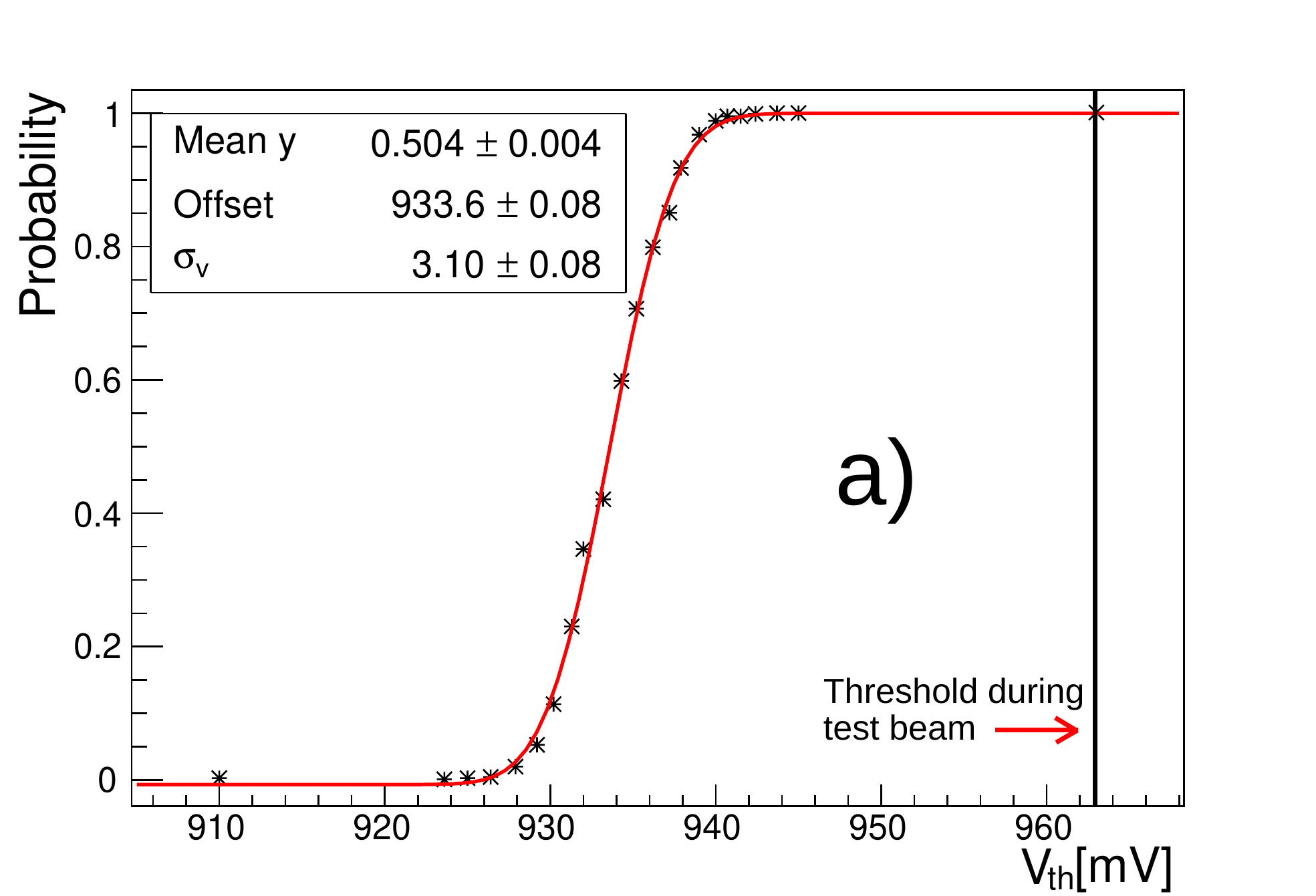}
\end{subfigure}
\begin{subfigure}{0.49\textwidth}
\centering
\includegraphics[width = 0.92\textwidth]{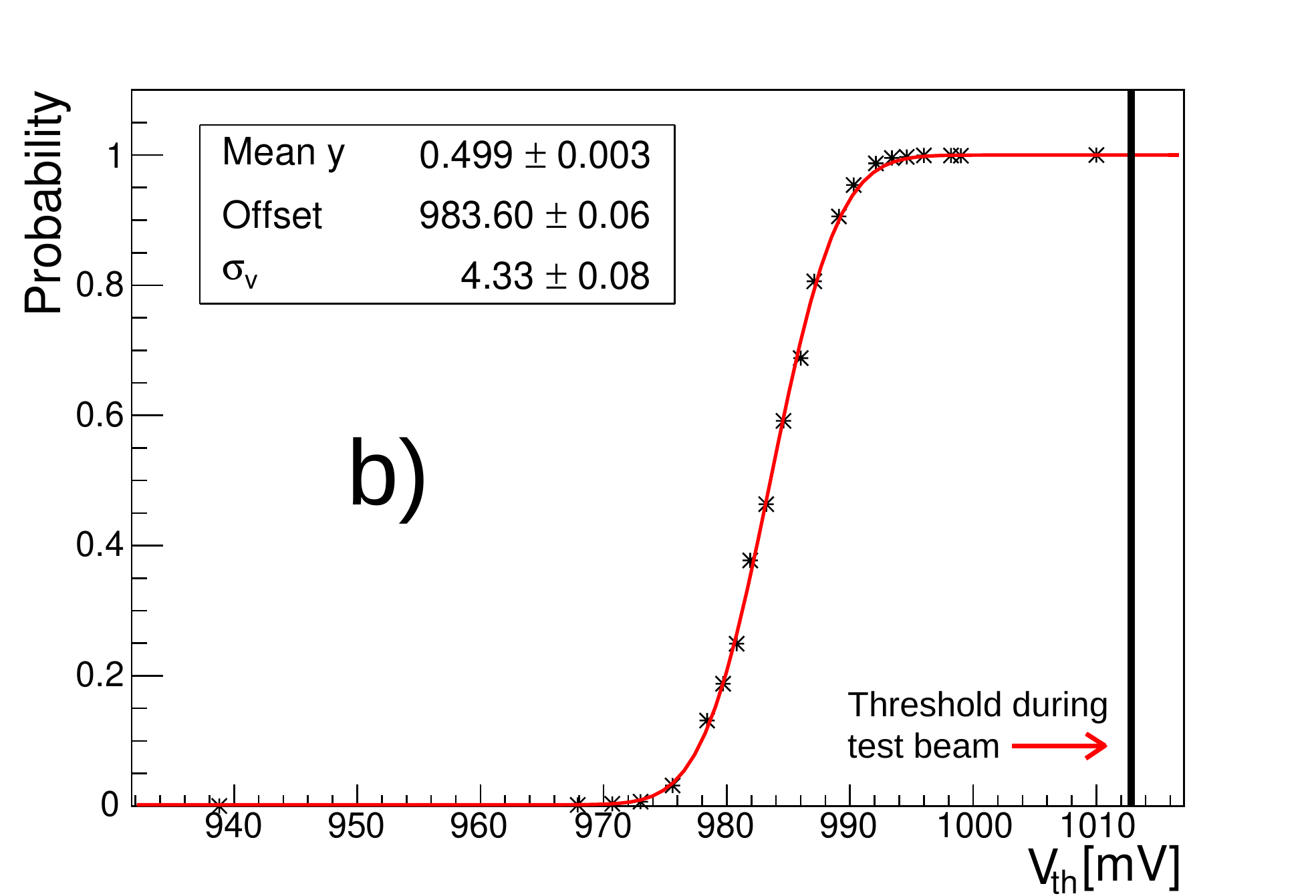}
\end{subfigure}
\caption{\label{fig:errorf} Average output state of the discriminator as a function of the threshold value for the upstream small pixel (a) and upstream large pixel (b). The curves were fitted with an error function; the output parameters of the fits are shown in the panels. The vertical lines show the value of the thresholds used during the test beam measurements}
\end{figure}

Device operation at $ 99 \% $ efficiency with charge sharing enables the estimation of the upper limit for the amplifier $ ENC $. Laboratory measurement of the error function of the small and large pixels, shown in Figure \ref{fig:errorf}, allow for the nominal threshold to be expressed as a function of voltage RMS noise of the amplifier, $ \sigma_{V} $. The thresholds for the small and large pixels are $ V^{small}_{th} = 9.5 ~\sigma^{small}_{V} $, and $ V^{large}_{th} = 6.8 ~\sigma^{large}_{V} $, respectively. Given a minimum primary charge $ Q_{MIN}=6250 ~\mathrm{e^{-}} $ from a MIP traversing $ 130 ~\mathrm{\mu m} $ of depleted silicon simulated by Geant4 \cite{GEANT4}, the upper limit for the $ ENC $ that allows operating at the plateau of efficiency with $ 15 \% $ charge sharing is: 
\begin{subequations}\label{eq:enc}
\begin{align}
\label{eq:enc:1}
ENC_{SMALL} \lesssim 0.85 \cdot \frac{Q_{MIN}}{9.5} = 560 ~\mathrm{e^{-}} ~\mathrm{RMS}
\\
\label{eq:enc:2}
ENC_{LARGE} \lesssim 0.85 \cdot \frac{Q_{MIN}}{6.8} = 780 ~\mathrm{e^{-}} ~\mathrm{RMS}
\end{align}
\end{subequations}
This result is compatible with expectations from the Cadence Spectre simulations.

\subsection{Time resolution}
\label{sec:time}

\begin{figure}[htpb]
\begin{subfigure}{0.49\textwidth}
\centering
\includegraphics[width = 0.92\textwidth]{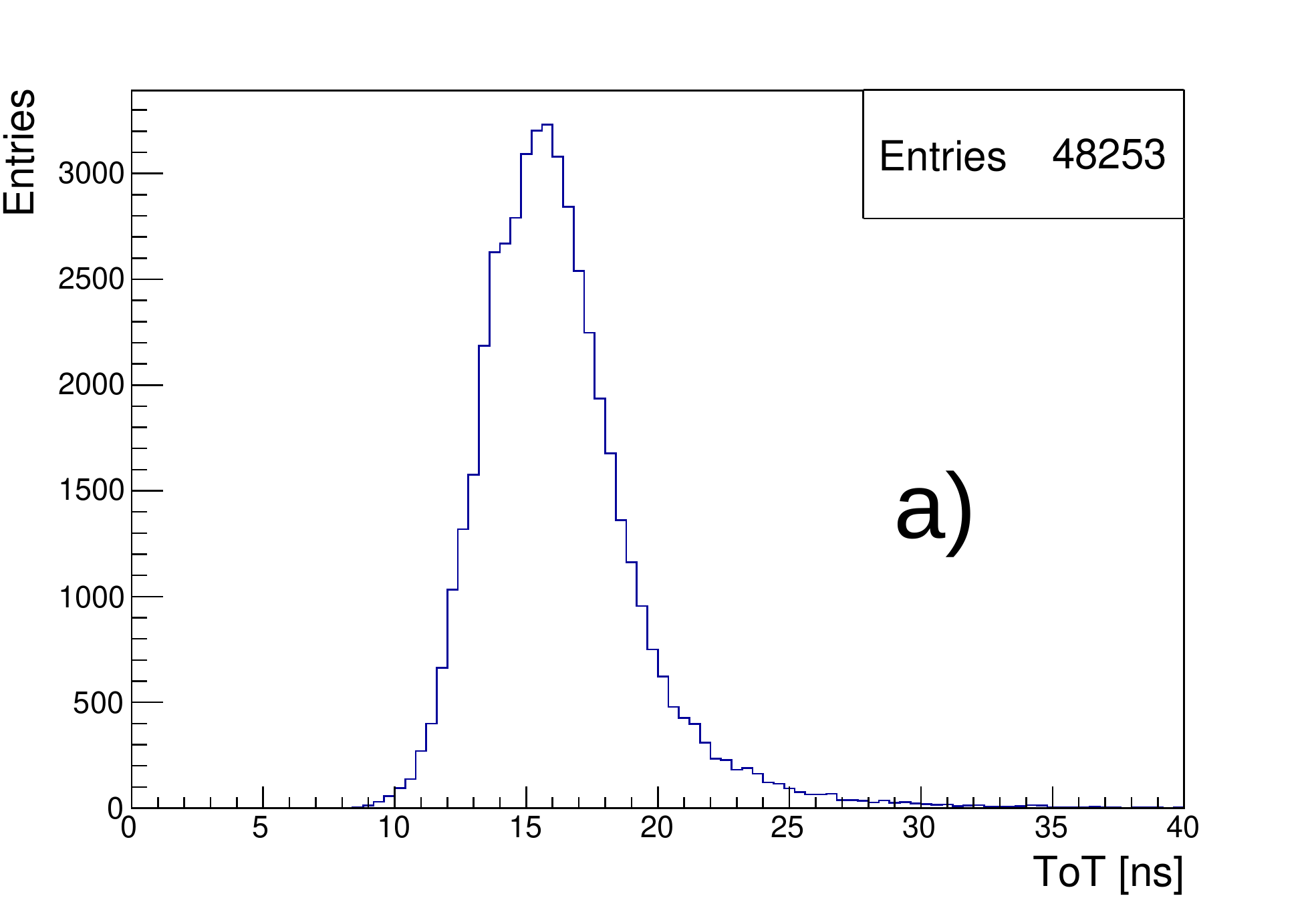}
\end{subfigure}
\begin{subfigure}{0.49\textwidth}
\centering
\includegraphics[width = 0.92\textwidth]{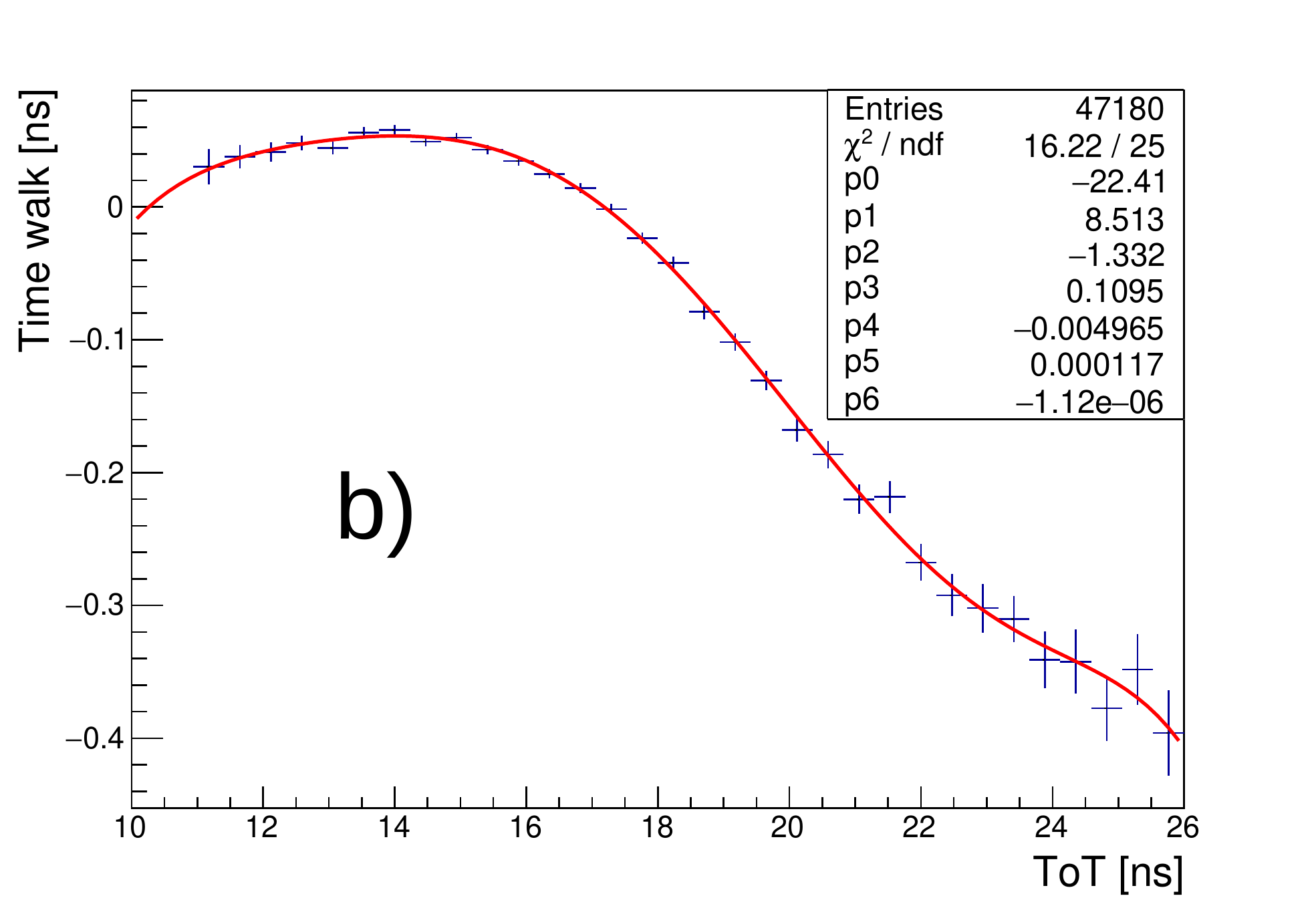}
\end{subfigure}
\caption{\label{fig:tw} a) Time-over-threshold distribution (a) and average time walk of the signal as a function of the time-over-threshold (b) for the small pixel of the upstream sensor. The vertical offset is arbitrary. The time walk correction is done by fitting the data with a polynomial function.}
\end{figure}

The time resolution of the three pixels under study has been measured in the region on interest described in Figure \ref{fig:EFFMAP}. Figure \ref{fig:tw} shows the ToT distribution and the time walk for the small pixel of the upstream sensor. The time walk correction is performed using the polynomial function resulting from the fit. The spread of the time walk distribution and the signal-to-noise ratio allows for the estimation of the signal peaking time at the output of the amplifier; we find it to be less than $ 2 ~\mathrm{ns} $, consistent with the simulation for the present configuration of the detector.

\begin{figure}[htpb]
\begin{subfigure}{0.49\textwidth}
\centering
\includegraphics[width = 0.92\textwidth]{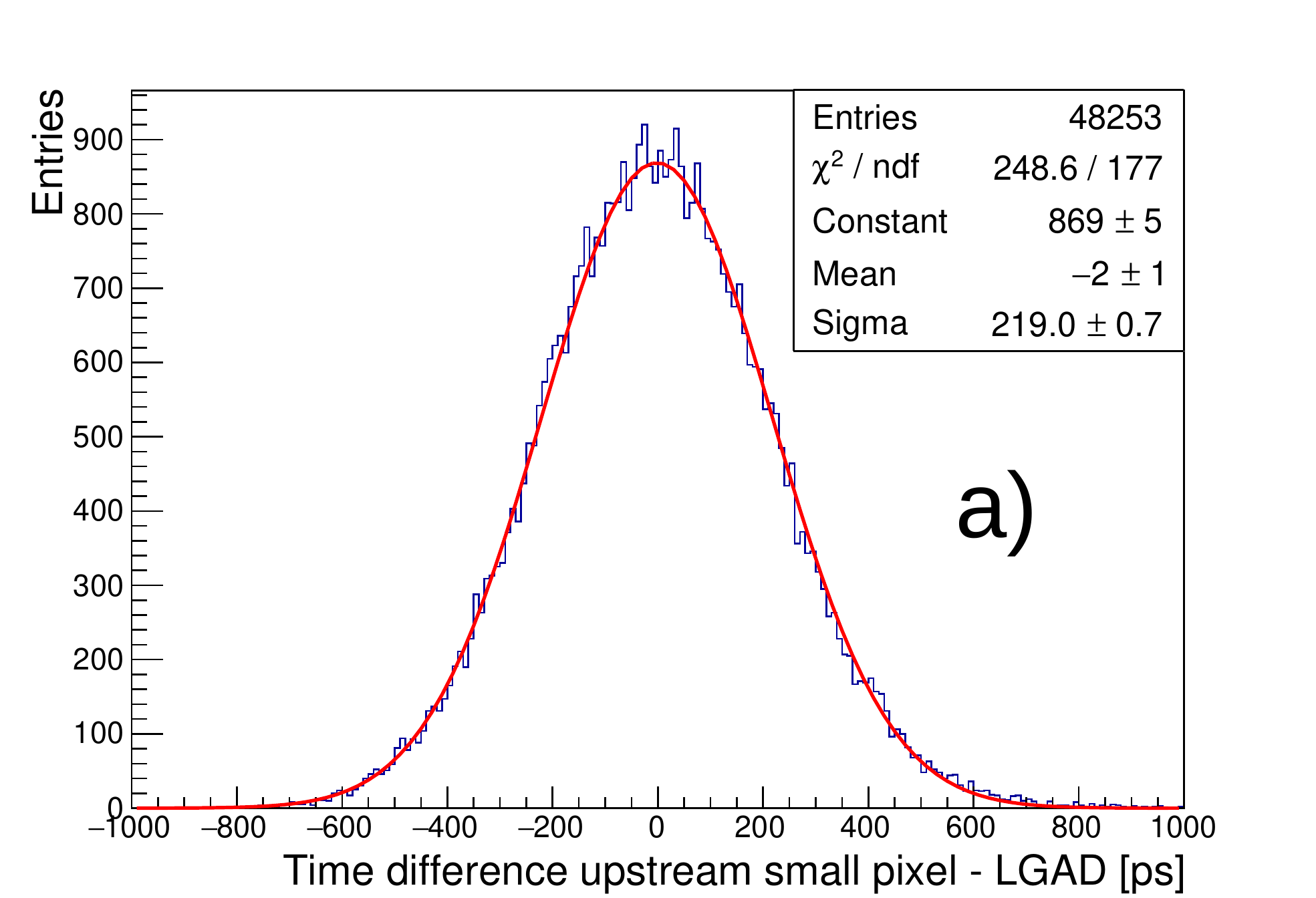}
\end{subfigure}
\begin{subfigure}{0.49\textwidth}
\centering
\includegraphics[width = 0.92\textwidth]{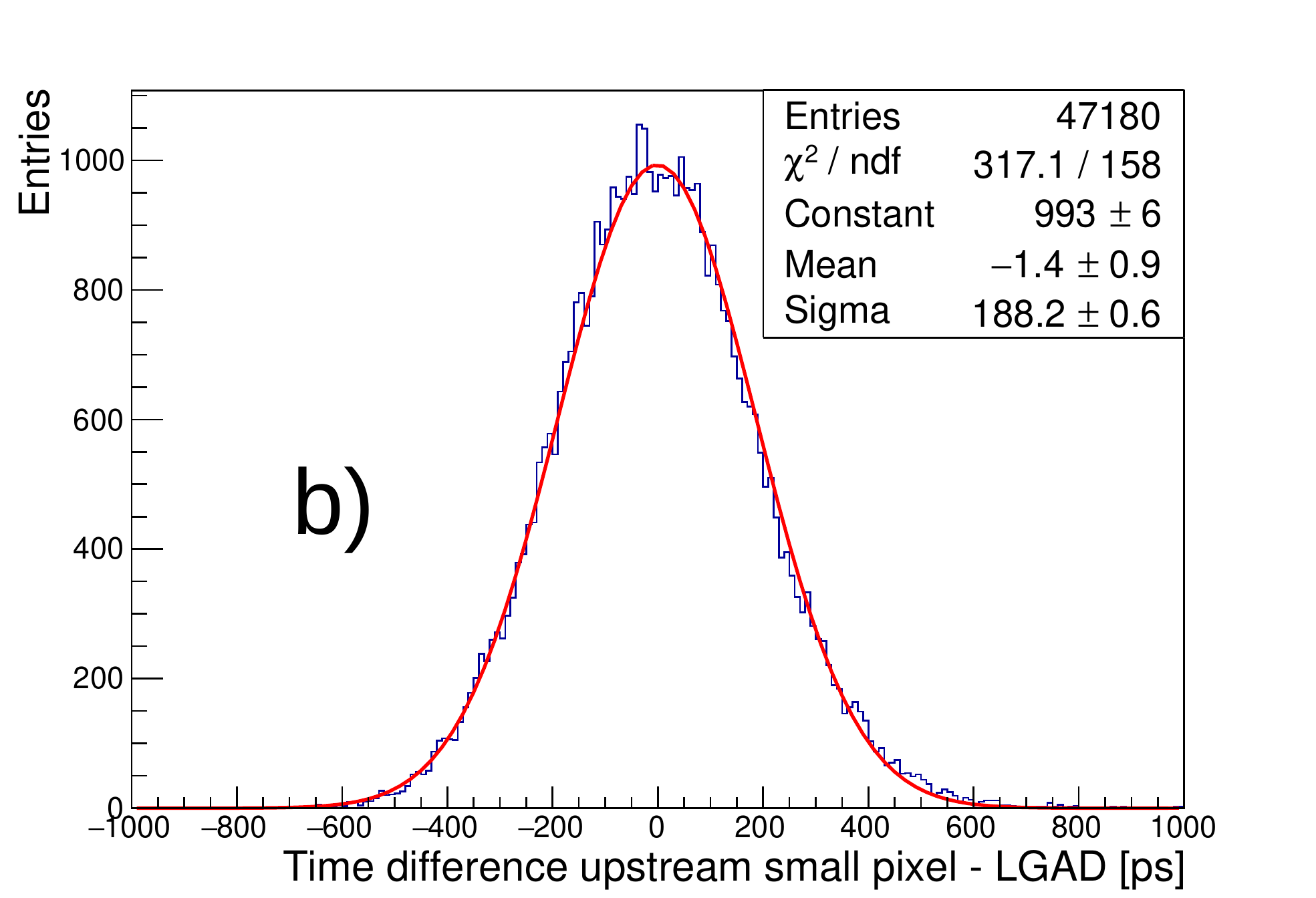}
\end{subfigure}
\caption{\label{fig:timeres} Measurement of the jitter on the time difference between the small pixel of the upstream monolithic sensor and the LGAD without correcting for the particle hit position (a) and correcting for it (b). The mean value of the time difference distributions is set to zero by the time walk correction. The data are fitted with a gaussian distribution to measure the standard deviation.}
\end{figure}

The time resolution after the time walk correction of the small pixel from the upstream sensor is shown in Figure \ref{fig:timeres} (a). This result is degraded by the non-uniformity of the response of the sensor due to the lack of backplane metallization. This effect can be partly corrected by measuring the variation of the average time-of-flight as a function of the particle hit position on the sensor surface. Figure \ref{fig:timeres} (b) shows the time resolution obtained for the same pixel after this correction. Following the same analysis procedure, the time resolution measured for the three pixels under study, before and after the correction for the particle hit position, is reported in Table \ref{tab:tres}.



\begin{table}[htbp]
\centering
\caption{\label{tab:tres} Measured time resolution for the monolithic sensor in SiGe BiCMOS technology for MIPs. The contribution from the LGAD sensor is neglected. The capacitance of the small and large pixels is $ 0.8 $ and $ 1.2 ~\mathrm{pF} $, respectively.}
\smallskip
\begin{tabular}{l|cc|}
\cline{2-3}
& \multicolumn{2}{ c| }{Time resolution $ \left[ \mathrm{ps} \right] $}\\
\hline
\multicolumn{1}{ |c|  }{Pixel} & w/o position correction & with position correction\\
\hline
\multicolumn{1}{ |l|  }{Downstream Small} & $ 202.3 ~\pm ~0.8 $ & $ 167.7 ~\pm ~0.7 $\\
\multicolumn{1}{ |l| }{Upstream Small} & $ 219.0 ~\pm ~0.7 $ & $ 188.2 ~\pm ~0.6 $\\
\hline
\multicolumn{1}{ |l|  }{Upstream Large} & $  265 ~\pm ~1 $ & $ 212 ~\pm ~1 $\\
\hline
\end{tabular}
\end{table}

The large pixel shows $ \sim 20\% $ worse time resolution with respect to the small pixels. Since the $ ENC $ is proportional to the pixel capacitance, the difference is the detector capacitance of a factor $ 1.5 $ between the large and the small pixels under study, should lead to an increase of the time resolution by the same factor. The smaller difference measured implies that the time resolution is affected, for both pixels, by an additional term, which is not proportional to the amplifier $ ENC $. This term could be attributed to the non-uniformity of the electric field, the thick substrate and the absence of the sensor backplane metallization.

\section{Conclusions}
\label{sec:conclusions}

A test beam measurement of the noise, efficiency, and time resolution of the first prototype of the TT-PET monolithic silicon pixel detector in the SiGe BiCMOS technology SG13S from IHP was performed at the CERN SPS with $ 180 ~\mathrm{GeV/c} $ pions. The detector, optimized for precise time measurement, has a preamplifier equivalent noise charge of less than $ 600 ~\mathrm{e^{-}} ~\mathrm{RMS} $ for a $ 0.8 ~\mathrm{pF} $ pixel capacitance, with a total time walk of less than $ 1 ~\mathrm{ns} $ for minimum-ionizing particle signals. This performance is obtained with a power consumption of $ 350 ~\mathrm{\mu W/channel} $. The detectors tested show an efficiency larger than $ 99\% $. Despite the absence of crucial chip processing steps, such as wafer thinning and backplane metallization, the time resolution is measured to be approximately $ 200 ~\mathrm{ps} ~\mathrm{RMS} $ for pixels with $ 0.8 ~\mathrm{pF} $ capacitance.

\acknowledgments

The authors wish to thank Florentina Manolescu from the CERN Bondlab for her excellent work, which made the test beam possible, Gabriel Pelleriti from the University of Geneva for the preparation of the detector board and the staff of SPS North Area at CERN for providing excellent quality beam for the whole duration of the test. A particular thank goes to Sebastian Grinstein and his colleagues for providing the LGAD sensor used as a time reference for this test. We are grateful to the colleagues of the University of Geneva who collaborated to the setup and operation of the test beam telescope. Finally, the authours wish to thank the Swiss National Science Foundation, which supported this research \cite{grant}.




\end{document}